\documentclass[prd,twocolumn,showpacs,floatfix,superscriptaddress]{revtex4}
\usepackage{graphicx}
\usepackage{epsfig}
\usepackage{bm}
\usepackage{amssymb}
\usepackage{float}
\usepackage{amsmath}
\usepackage{dcolumn}
\usepackage{cancel}
\usepackage[colorlinks]{hyperref}
\usepackage[usenames,dvipsnames]{color}
\hypersetup{
     breaklinks=true,
    pdfstartview={FitH},    
    colorlinks=true,       
    linkcolor=blue,          
    citecolor=red,        
    filecolor=magenta,      
    urlcolor=blue,           
    anchorcolor=green,      
    linktocpage=true
}

\def\doi{http://doi.org}
\def\arxiv{http://arxiv.org/abs}
 
 \def\e{\mathrm{e}}

\begin{document}

\title{Holographic bounce}

 \author{Shin'ichi Nojiri}
\email{nojiri@gravity.phys.nagoya-u.ac.jp}
\affiliation{Department of Physics, Nagoya University, Nagoya 464-8602, Japan}
\affiliation{Kobayashi-Maskawa Institute for the Origin of Particles and the
Universe, Nagoya University, Nagoya 464-8602, Japan}

\author{Sergei D. Odintsov}
\email{odintsov@ieec.uab.es}
\affiliation{Institut de Ciencies de lEspai (IEEC-CSIC), 
Campus UAB, Carrer de Can Magrans, s/n 
08193 Cerdanyola del Valles, Barcelona, Spain}
\affiliation{Instituci\'{o} Catalana de Recerca i Estudis Avan\c{c}ats
(ICREA), Passeig Llu\'{i}s Companys, 23 08010 Barcelona, Spain}

\author{Emmanuel N. Saridakis}
\email{msaridak@phys.uoa.gr}
\affiliation{Department of Physics, National Technical University of Athens, Zografou
Campus GR 157 73, Athens, Greece}
\affiliation{Department of Astronomy, School of Physical Sciences, University of Science 
and Technology of China, Hefei 230026, P.R. China}
\affiliation{Chongqing University of Posts \& Telecommunications, Chongqing, 
400065, P.R.
China}

\begin{abstract} 
We investigate the bounce realization arising from the application of the holographic 
principle in the early universe, inspired by  its  well-studied 
late-time application. We first consider as Infrared cutoffs the particle and future 
event horizons, and we show that  the decrease of the horizons at 
early times naturally increases holographic energy density at bouncing scales, while we 
additionally obtain the necessary null energy condition 
violation. Furthermore, adding a simple correction to the horizons due to the 
Ultraviolet cutoff we analytically obtain improved nonsingular bouncing solutions, in 
which the   value of the  minimum scale factor is controlled by the UV correction.
Finally, we  construct generalized scenarios, arisen 
from the use of  extended Infrared cutoffs, and as specific examples we consider 
cutoffs that can reproduce $F(R)$ gravity, and the bounce realization within it.

\end{abstract}


\maketitle

\section{Introduction}

One of the interesting explanations for the late-time acceleration is the holographic 
dark energy scenario \cite{Li:2004rb} (for a review see \cite{Wang:2016och}). In this 
framework the dark energy sector arises  holographically from the vacuum energy, after one 
applies at a cosmological level the  holographic 
principle, which originating from black hole thermodynamics and string theory 
\cite{tHooft:1993dmi,Susskind:1994vu,Witten:1998qj,Bousso:2002ju,Kiritsis:2013gia}   
establishes a 
connection  of the  largest distance of this theory (related to 
causality and the  quantum field theory  applicability at large distances)
\cite{Cohen:1998zx} with the Ultraviolet cutoff of a quantum field theory (related to 
the vacuum energy). 
Holographic dark energy, both in 
its basic 
\cite{Li:2004rb,Wang:2016och,Horvat:2004vn,Huang:2004ai,Enqvist:2004xv,Zhang:2005yz,
Elizalde:2005ju,Pavon:2005yx, Guberina:2005fb, Wang:2005jx, 
Nojiri:2005pu,Setare:2008pc} as well 
as in its various extensions 
\cite{Gong:2004fq,Ito:2004qi,Gong:2004cb,Setare:2008bb,Saridakis:2007ns,
Gong:2009dc,BouhmadiLopez:2011xi,Khurshudyan:2014axa,Pasqua:2016wrm,
Jawad:2016tne,Pourhassan:2017cba,Paul:2019hys},
proves to be very efficient in describing the late-time universe, and it is in agreement 
with
observations
\cite{Huang:2004wt,Wang:2004nqa,Zhang:2005hs,Li:2009bn, Lu:2009iv}.

Although there has been a large amount of research in the late-time application of 
holographic principle, namely in the dark-energy era, there has been almost no effort in 
applying it at early times. Recently, we proceeded to such direction and we constructed 
  an inflationary realization of holographic origin \cite{Nojiri:2019kkp}. The scenario 
has the advantage that  since at early times the 
largest distance of the theory  is small, the holographic energy density is naturally 
large in order to lie in the regime necessary for inflation.

In the present work we are interested in applying the holographic principle at early 
times, however investigating the bounce realization. Nonsingular bouncing cosmologies are 
of significant interest since they offer a potential solution to the 
cosmological singularity problem and thus an alternative to inflation  
\cite{Novello:2008ra,Cai:2012va, Brandenberger:2012zb,Cai:2014bea, 
 Brandenberger:2016vhg, Cai:2016hea,deHaro:2015wda}. In order for the bounce to be 
realized the null energy condition has to be violated, and this can be 
obtained by various 
modified 
gravity constructions \cite{Nojiri:2006ri, Capozziello:2011et, Cai:2015emx}, such as the 
Pre-Big-Bang \cite{Veneziano:1991ek} and the Ekpyrotic \cite{Khoury:2001wf} 
scenarios,  $f(R)$ gravity 
\cite{Bamba:2013fha,  Pavlovic:2017umo}, $f(T)$ 
gravity \cite{Cai:2011tc},      loop 
quantum cosmology \cite{Bojowald:2001xe, Cai:2014zga, 
Odintsov:2015uca}, Lagrange modified gravity \cite{Cai:2010zma}, Finsler gravity 
\cite{Minas:2019urp} etc.
Alternatively, a
nonsingular bounce  may be acquired through the   introduction of exotic matter sectors 
 \cite{Cai:2007qw,   Cai:2009zp}.
Indeed, as we will see in this work, the energy 
density of holographic origin can lead to the necessary violation of the null energy 
condition and trigger the bounce. Additionally,   the decrease of the horizons at early 
times naturally 
increases holographic energy density at bouncing scales.

The plan of the work is the following. In Section \ref{model} we construct the 
holographic bounce scenario, extracting the corresponding analytical solutions. In 
Section \ref{Generalizedmodel} we proceed to the construction of generalized scenarios, 
with 
extended Infrared cutoffs. Finally, Section \ref{Conclusions} is devoted to the 
conclusions.

\section{Holographic bounce}
\label{model}

In this section we will  construct the basic model of holographic bounce. We start by 
reminding that in general the holographic  density is proportional to 
the inverse squared Infrared cutoff $ L_\mathrm{IR}$, namely 
 \begin{equation}
 \label{basicnew}
\rho=\frac{3c^2}{\kappa^2 L^2_\mathrm{IR}},
\end{equation}
where $\kappa^2$ is the gravitational constant and $c$ a parameter. 
If one wishes to apply the above relation in a cosmological framework, he considers the 
homogeneous and isotropic Friedmann-Robertson-Walker (FRW)
  metric
\begin{equation}
\label{FRWmetricnew}
ds^2=- dt^2+a^2(t) \left(\frac{dr^2}{1-kr^2}+r^2 d\Omega^2\right),
\end{equation}
with $a(t)$   the scale factor and  
$k=0,+1,-1$ corresponding  to flat, close and open spatial geometry respectively. In the 
following we focus on the flat case, however the 
generalization to non-flat geometry is straightforward.

Since the Infrared cutoff $ L_\mathrm{IR}$ is related to causality it must be 
a form of horizon. The simplest choice is the Hubble radius, however concerning the 
late-time application its use cannot lead to an accelerating universe and thus it is 
rejected \cite{Hsu:2004ri}. Hence, one may use  
the particle and future event horizons \cite{Li:2004rb}, the age of the universe  
 \cite{Cai:2007us,Wei:2007ty}, the 
inverse square root of the Ricci curvature
\cite{Gao:2007ep}, or a combination of Ricci and Gauss-Bonnet invariants 
\cite{Saridakis:2017rdo}. More generally, one can consider a   
general  Infrared cutoff constructed by an arbitrary combination of all the above 
quantities and their derivatives
 \cite{Nojiri:2017opc}. In this Section, as a first application we will consider 
 the particle horizon $L_p$ 
and the future event horizon $L_f$, which are written as
\begin{equation}
\label{LpLf}
L_p\equiv a
\int_0^t\frac{dt}{a}\ ,\quad L_f\equiv a \int_t^\infty \frac{dt}{a}\, .
\end{equation}
 
Although in the late universe the Friedmann equation includes both the matter and 
holographic dark energy sectors, in the case of the early universe the former can be 
neglected \cite{Nojiri:2019kkp} (unless one considers the possibility of a ``matter 
bounce'' \cite{Brandenberger:2012zb,Cai:2011tc}, which is not the case of the present 
work). Therefore, at early times the first Friedmann equation is written 
as
\begin{equation}\label{FR1}
H^2=\frac{\kappa^2}{3} \rho,
\end{equation}
where $\rho$ is the  energy density of the (effective) fluid that consists the 
universe, originating from a scalar field, from modified-gravity, or from 
other sources. In this manuscript we consider that the   fluid that drives the 
bounce   has a holographic origin, i.e. it arises from the holographic energy density. 
Thus, imposing that  
$\rho$ in (\ref{FR1})  is $\rho$ of (\ref{basicnew}) we deduce that 
\begin{equation}
\label{H2}
H^2=\frac{c^2}{L^2_\mathrm{IR}}\, .
\end{equation} 
Hence, inserting the   particle horizon $L_p$ 
or the future event horizon $L_f$ into   (\ref{H2}) we find that  
\begin{equation}
\label{H5new}
\frac{d}{dt}\left(\frac{1}{a}\sqrt{\frac{c^2}{H^2}}\right)= \frac{m}{a}\, ,
\end{equation}
with $m=1$  corresponding to the particle horizon and $m=-1$ to the future event 
horizon. 
 
The general solution of equation (\ref{H5new}) is
\begin{equation}
\label{sol1}
a(t)=a_0 \left(t-t_0\right)^{\frac{c}{c\pm m}},
\end{equation}
with $a_0$,$t_0$ the two integration constants and $\pm$ corresponding to the two 
solution branches. As we observe the above solution has very interesting expressions for 
particular values of the parameter $c$. Specifically, in the case where $\frac{c}{c\pm 
m}$ is an even number we obtain a bouncing scale factor, which is the scope of interest 
of the present work. This can be obtained  taking the $+$ branch in the case of $m=-1$, 
i.e. using  the future event 
horizon, and taking the $-$ branch in the case where  $m=1$, i.e. using the particle 
horizon. For instance, choosing $c=2$ we can see that in both the above cases we obtain 
 $a(t)=a_0 \left(t-t_0\right)^2$ which indeed corresponds to the bounce realization. 
 In summary, as we observe, in the scenario at hand the   bounce  can be 
straightforwardly obtained, since  the decrease of the horizons at early 
times naturally increases holographic energy density at bouncing scales.

We proceed by investigating an improved version of the above scenario. In particular, 
since we apply the holographic principle at early times, i.e. at high energy scales, we 
should incorporate the Ultraviolet cutoff $\Lambda_\mathrm{UV}$ too. Namely, in this 
regime 
the Infrared cutoff acquires a correction from the 
Ultraviolet one, which takes the   form 
\cite{Nojiri:2004pf} 
\begin{equation}
\label{H8cnew}
L_\mathrm{IR} \rightarrow \sqrt{  L^2_\mathrm{IR} + \frac{1}{\Lambda_\mathrm{UV}^2}}\, .
\end{equation}
If we insert this  expression into (\ref{H2}), with $L_\mathrm{IR}$
  either the    particle horizon $L_p$ 
or the future event horizon $L_f$, we acquire 
 \begin{equation}
\label{H8d}
\frac{m}{a} =  \frac{d}{dt} \left( \frac{1}{a} \sqrt{ \frac{c^2}{H^2} 
 - \frac{1}{\Lambda_\mathrm{UV}^2}} \right) \, ,
\end{equation}
which is the improved version of  (\ref{H5new}). In order to solve this equation we first 
transform it to an equation for $H(t)$, namely 
\begin{equation}
\label{H8enew}
\dot H = - \frac{H^3}{c^2} \left\{m \sqrt{ \frac{c^2}{H^2}  - 
\frac{1}{\Lambda_\mathrm{UV}^2}} 
+ H \left( \frac{c^2}{H^2} - \frac{1}{\Lambda_\mathrm{UV}^2} \right) \right\} \, ,
\end{equation}
 whose general solution  for $\Lambda_\mathrm{UV}$ not being equal 
to 0 or infinity is written   implicitly as
{\small{\begin{eqnarray}
&&
\!\!\!\!\!\!\!\!\!\!\!\!\!
\frac{\sqrt{c^2-\frac{H^2}{\Lambda_\mathrm{UV}^2}}}{
(c^2-1)^{\frac{3}{2}} \Lambda_\mathrm{UV}^2 \sqrt{H^2-c^2\Lambda_\mathrm{UV}^2}}
\tan^{-1}\left[
\frac{mH}{\sqrt{c^2-1} \sqrt{H^2-c^2\Lambda_\mathrm{UV}^2}  } \right]
\nonumber\\
&&
\!\!\!\!\!\!\!\!\!\!\!\!\!
+\frac{m\sqrt{c^2\!-\!\frac{H^2}{\Lambda_\mathrm{UV}^2}}\!-\!c^2}{
c^2(c^2\!-\!1)H\Lambda_\mathrm{UV} ^2}+
\frac{
\tanh^{-1}\!\left[
\frac{H}{\sqrt{c^2\!-\!1}\Lambda_\mathrm{UV}} \right]}
{ (c^2-1)^{\frac{3}{2}}\Lambda_\mathrm{UV}^3 }
 =\!-\!\frac{t}{c^2 \Lambda_\mathrm{UV}^2}\!+\!C_0,
 \label{generalsolution}
\end{eqnarray}}}
with $C_0$ an integration constant. Hence, the scale factor can be obtained through 
\begin{equation}
\label{at}
a(t)= a_0 e^{\int H(t)dt},
\end{equation}
with $a_0$ another integration constant.
 
\begin{figure}[ht]
\centering
\includegraphics[scale=.40]{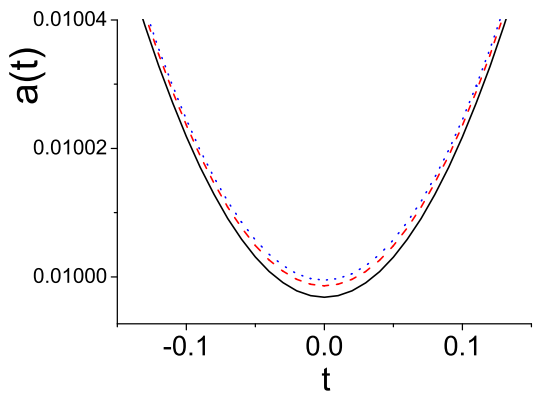}
\caption{{\it{ The scale factor evolution in the holographic bounce scenario arising from 
the analytical solution (\ref{generalsolution}), for 
 $m=-1$ (i.e. when the future event horizon is used),
 with 
$c=2$, $\Lambda_\mathrm{UV}=20$ (black-solid curve),
$c=2$, $\Lambda_\mathrm{UV}=30$ (red-dashed curve),
$c=2$, $\Lambda_\mathrm{UV}=50$ (blue-dotted curve), in units where $\kappa^2 = 1$, 
setting the bouncing point at $t=0$ and $a(t=-10)=0.25$.
}}}
\label{bounce}
\end{figure}

The above solution (\ref{generalsolution}) can describe a bounce for particular choices 
of the model parameters that satisfy the bounce requirements, namely $H<0$, $H=0$ and 
$H>0$ before, at, and after the bounce respectively, as well as $\dot{H}>0$ throughout 
the procedure. In Fig. \ref{bounce} we present the bounce realization arising from  
(\ref{generalsolution}), for three choices of the model parameters. The interesting 
feature is that now we can obtain a nonsingular bounce, which was not possible in the 
simple case without the UV correction, namely solution (\ref{sol1}). Furthermore, as we 
can see, the value of the minimum scale factor is determined by the value of  
$\Lambda_\mathrm{UV}$, going to zero at $\Lambda_\mathrm{UV}\rightarrow\infty$ as 
expected. This feature is very significant, since nonsingular bounces are more physically 
important since they are the ones that can alleviate the initial singularity issue of 
standard cosmology. This is one of the main result of the present work 
and reveals the capabilities of holographic bounce.

We proceed by referring to the perturbations of the above background 
evolution, which is a necessary task in every bouncing scenario, since they are related to 
observables such as the spectral index and the tensor-to-scalar ratio. In particular, 
while in inflation  the cosmological fluctuations emerge initially  inside the Hubble 
horizon, then they exit it, and later on they  re-enter giving rise to the   primordial 
power spectrum, in   bounce cosmology the situation is different. Namely,   the quantum
fluctuations around the initial vacuum state are generated in the contracting phase 
before the bounce, and since  the 
Hubble horizon decreases faster than the wavelengths of the primordial fluctuations, 
they exit the Hubble radius  as contraction continues. Then the bounce happens, and hence  
at later times in the expanding phase the fluctuations  re-enter inside the horizon. As 
one can deduce, the specific background evolution could affect the processing of 
perturbations from the bounce point, however  this effect is important only in the 
UV regime, where the gravitational modification becomes significant,
while the IR regime, which is responsible for  the   primordial perturbations, remains 
almost unaffected 
\cite{Battarra:2014tga,Quintin:2015rta}.

Although the generation of perturbations in bouncing models driven by a scalar 
degree of freedom is well understood and studied 
\cite{Novello:2008ra,Allen:2004vz}, in the case where the driving mechanism is 
different the perturbation generation has not been studied in detail. Therefore,  
the analysis of perturbation 
generation in the present holographic bounce scenario   has to be performed in a  
systematic way, through the full and detailed perturbation   analysis, using the 
perturbation investigation of holographic cosmology 
\cite{Battefeld:2004jh,Kim:2008pi,Mehrabi:2015kta}. Only such an analysis could  
reveal whether the present bouncing scenario suffers from the ``observational no-go
theorem'' that widely exists in nonsingular bounces, namely that their predicted   
amplitude of primordial non-gaussianity and  
tensor-to-scalar ratio cannot simultaneously fit observations in a
general single component bounce cosmology \cite{Quintin:2015rta, Li:2016xjb} (the fact 
hat holographic bounce is not driven by a single component is a promising feature that 
the above no-go theorem might be evaded). Furthermore, the perturbation analysis would 
reveal the differences of the present scenario form other scenarios, such as 
the ekpyrotic  \cite{Khoury:2001wf} and the matter bounce \cite{Cai:2011tc} ones, since 
even if the background behavior is similar, the perturbation evolution will be different 
due to the fact that the underlying theory and the propagating degrees of freedom are of 
different, namely of holographic, origin.
Nevertheless, this detailed analysis of perturbation generation  is a separate 
project  that lies beyond the scope of the present work, and it is 
left for a future study.

One issue that is present in principle in almost every bouncing scenario, is the 
anisotropy one. Specifically, the bounce 
realization is in general unstable against  anisotropic stress  (BKL 
instability) \cite{Belinsky:1970ew}, since   the effective energy density in anisotropies 
scales as  $a(t)^{-6}$ and thus in a contracting 
universe it increases   faster than the radiation  and matter energy densities, leading 
the   bounce to take place in a non-isotropic and non-homogeneous 
spacetime. Nevertheless,  although the anisotropies grow faster than the background 
evolution, their domination is not necessarily quick,  since this is determined by the   
initial conditions on the anisotropy. Hence, if the anisotropies are of quantum origin, 
arising from the backreaction of cosmological 
perturbations,   the moment where they will 
dominate  will depend on the energy scale during the matter 
contraction, which is typically at very high energy scales. Therefore, one may 
consider reliably  an FRW background evolution up to the bounce phase.
However, in order 
to acquire a full removal of the anisotropy problem, one should introduce additional 
mechanisms (a well-studied such scenario is the ekpyrotic one \cite{Khoury:2001wf}, in 
which one  introduces  a negative exponential potential that leads to an anisotropy-free 
evolution \cite{Cai:2012va,Cai:2014bea,Qiu:2013eoa,Cai:2013vm}). Nevertheless, since in 
the present scenario of holographic bounce  there is no extra degree of freedom, such 
mechanisms are difficult to be introduced, ad thus a safe removal of the anisotropy 
problem should be obtained by imposing the bounce scale to be suitably low, in order for 
the anisotropies to be always sub-dominant.

Finally, let us make a comment on the preheating phase, which is is needed in any 
realistic cosmological scenario. The preheating phase in bouncing cosmology has been 
investigated in \cite{Cai:2011ci}. Nevertheless, it is interesting to note that in the 
holographic framework there can be a mechanism of ``holographic preheating'' 
\cite{Cai:2016sdu,Cai:2016lqa}, through which 
the energy transfer from the inflaton to the matter field is obtained through  a model of 
holographic superconductor. Hence, due to the holographic foundations of the present 
scenario, the phase of holographic preheating can be naturally embedded to it.

\section{Generalized scenarios} 
\label{Generalizedmodel}

In this section we construct generalizations of the above holographic bounce scenarios 
through the use of extended Infrared cutoffs, inspired by similar constructions in the 
case of the  late universe. In principle, such general Infrared cut-offs
may be a function of both   $L_p$ and $L_f$   and their derivatives, as well as 
of the Hubble horizon and its derivatives and of the scale factor  
\cite{Nojiri:2005pu,Nojiri:2017opc}, namely
\begin{equation}
\label{generalLIR}\!\!\!
L_\mathrm{IR}\!=\!L_\mathrm{IR}\!\left(L_p, \dot L_p, \ddot L_p, \cdots\!, L_f, \dot L_f, 
\ddot 
L_f, \cdots\!, a,
H, \dot H, \ddot H, \cdots\! \right)\! .
 \end{equation}
Thus, applying the above  extended Infrared cutoff at the early universe provides 
enhanced freedom to acquire  bouncing cosmology.

Without loss of generality we start by considering the model 
\begin{equation}
\label{H10}
L_\mathrm{IR}= \frac{1}{6\alpha{\dot H}^2 a^6}  \int^t dt a^{6} \dot H \, ,
\end{equation}
with  $\alpha$ a parameter. Inserting this into the Friedmann equation (\ref{H2}), 
setting  $c=1$ for simplicity, and adding also a cosmological constant term $\Lambda = - 
\frac{3}{\kappa^2} H_0^2$, we result to 
 \begin{equation}
\label{HBC10}
0 = - 3 \left( H^2 - H_0^2 \right) 
 - \alpha \left( - 108 H^2 \dot H + 18 {\dot H}^2 - 36 H \ddot H \right) \, .
\end{equation}
Recalling that  the Ricci scalar is   
$R=6(2H^2+\dot{H})$ in FRW geometry, the 
above   equation is written in the form 
\begin{equation}
\label{H12}
0 = \frac{F(R)}{2} - 3(H^2\! +\! \dot H) F'(R)
 + 18 ( 4H^2 \dot H\! +\! H \ddot H) F''(R)\, ,
\end{equation}
with 
\begin{equation}
\label{HBC11}
F(R) = R - \alpha R^2 - 6 H_0^2\, .
\end{equation}
As one can see, 
Eq.(\ref{H12}) is just   the first Friedmann equation in  the $R^2$-gravity theory in the 
absence of matter  \cite{Nojiri:2017ncd,Nojiri:2010wj,Nojiri:2006ri,
Capozziello:2010zz,delaCruzDombriz:2012xy,Olmo:2011uz}.

 As it is known, for the $F(R)$ gravity model of (\ref{HBC11}) there appears a solution 
describing the bouncing universe 
\cite{Bamba:2013fha}. 
In fact, if we impose the bouncing scale factor
\begin{equation}
\label{HBC12}
a(t) =a_B \e^{\frac{\lambda}{2} t^2} \, ,
\end{equation}
with a positive constant $\lambda$,
we find that $
H = \lambda t $, $ \dot H = \lambda $, and $\ddot H = 0 $ (i.e. the universe is 
contracting for $t<0$, is expanding for $t>0$, and at $t=0$ we have the nonsingular 
bounce realization at a scale factor $a_B$).
In this case 
Eq.~(\ref{HBC10}) becomes
\begin{equation}
\label{HBC14}
  \left( H_0^2 - 6 \alpha \lambda^2 \right)
-  \left( 1 - 36 \alpha \lambda \right) \lambda^2 t^2 
=0 \, ,
\end{equation}
which for  $\alpha>0$ is satisfied 
if we choose 
\begin{eqnarray}
&&H_0^2 = \frac{1}{216 \alpha}\, \nonumber\\
&&\lambda = \frac{1}{36\alpha} = 6 H_0^2\, .
\label{HBC15}
\end{eqnarray} 
Hence, the scenario at hand indeed accepts the bouncing scale factor (\ref{HBC12}) as 
solution, if we consider the model parameters  (\ref{HBC15}).
 We mention here that the scale factor (\ref{HBC12}) is just a simple example that 
shows the capabilities of the theory. In a realistic inflation realization however the 
imposed scale factor should include an exit from the exponential expansion. 
Alternatively, one could transform the above $F(R)$ model in the Einstein frame, where 
the scalar field dynamics in the effective potential could drive the successful inflation 
exit. In this way we can re-obtain the usual successful inflationary scenarios of $F(R)$ 
gravity, nevertheless the underlying theory is of different origin, namely it arises from 
holography.

As another example, we may consider the   model where
\begin{equation}
\label{HBC7}
L_\mathrm{IR} = \int^t dt \left( \frac{1}{\beta} - \frac{2 \beta B 
a(t)^{- \frac{1}{\beta}}}{H(t)^2} \right) \, ,
\end{equation}
with $\beta$ and $B$ two parameters. Inserting this  into the Friedmann equation 
(\ref{H2}), and setting for simplicity   $c=1$, we find  
\begin{equation}
\label{HBC4}
\dot H = 2 \beta B a^{- \frac{1}{\beta}} - \frac{1}{\beta} H^2 \, .
\end{equation}
The solution of (\ref{HBC4}) is 
\begin{equation}
\label{HBC2}
H = \frac{2 \beta B t}{A + B t^2} \, , 
\end{equation}
which in turn leads to  
 \begin{equation}
\label{HBC1}
a(t) = a_B\left( A + B t^2 \right)^\beta \, , 
\end{equation}
with   $A$ and $a_B$   integration constants. Indeed, the solution (\ref{HBC1}) describes 
a bounce realization at $t=0$. Moreover,  away from the bounce point
$a(t)$ behaves in a power-law way, namely $a(t) \sim t^{2\beta}$,  as in the case of the 
perfect fluid with a constant equation-of-state parameter.
Note that since 
for the solution (\ref{HBC1}) the integrand of (\ref{HBC7}) 
 diverges at $t=0$ and goes to a constant $\frac{1}{2\beta}$ when $t\to \infty$, it 
is necessary to add a small UV correction, as we discussed in the previous section, and 
use
\begin{equation}
\label{HBC18}
 L_\mathrm{IR} \rightarrow \frac{1}{2\beta}\int_0^t dt  - \int_t^\infty dt 
\left( \frac{1}{2\beta} 
 - \frac{2 \beta B a(t)^{- \frac{1}{\beta}}}{H(t)^2} \right)   \, .
\end{equation}
 
Finally, as a last example we  consider the model with
\begin{equation}
\label{HBC19}
L_\mathrm{IR} = \frac{1}{2\beta} \int^t dt 
\left(1 - \frac{A}{ a^{1/\beta} - A} \right)\, ,
\end{equation}
with $\beta$ and $A$   constants. 
inserting into (\ref{H2}) with $c=1$ we obtain  
\begin{equation}
\label{HBC20}
\dot H = \frac{H^2}{2\beta} \left(\frac{A}{ a^{1/\beta} - A} - 1 \right) \, ,
\end{equation}
whose solution is again given by (\ref{HBC1}) but now $B$ appears as 
a constant of   integration. 
Since substituting the solution into (\ref{HBC19}) we find that 
the second term in the integrand diverges at the bouncing time,
 the integration in (\ref{HBC19}) should be modified as 
\begin{equation}
\label{HBC22}
L_\mathrm{IR} \rightarrow \frac{1}{2\beta} \left( \int_0^t dt 
+ \int_t^\infty dt \frac{A}{ a^{1/\beta} - A} \right)\, ,
\end{equation}
i.e. we add a UV correction. \\ \\

\section{Conclusions} 
\label{Conclusions} 

In this work we obtained a bounce realization of holographic origin. In particular, 
inspired by the application of the holographic principle at late-time universe, i.e. 
by the scenario of holographic dark energy, we applied it at early times. We first  
considered as Infrared cutoffs the particle or future event horizons, and we showed that 
 the decrease of the horizons at early times naturally increases holographic energy 
density at bouncing scales, while we additionally obtain the necessary null energy 
condition 
violation. Hence, even this simple scenario can lead to the bounce realization. 

We proceeded to the  addition of  a simple correction due to the Ultraviolet cutoff, whose 
role may be taken into account at the high energy scales of early universe. We first
extracted analytical solutions that may lead to the bounce realization for suitable 
choices of the model parameters, and  we provided specific numerical examples.
The interesting implication of this UV correction is that it can control the value of 
the  minimum scale factor, which is significant since nonsingular 
bounces are more physically important relating to the alleviation of the  
initial singularity issue of 
standard cosmology.

Lastly, we   constructed generalized scenarios of holographic bouncing cosmology, arisen 
from the use of  extended Infrared cutoffs. As specific examples we considered cutoffs 
that can reproduce $F(R)$ gravity, and in particular  $R^2$ gravity, which can then lead 
to a nonsingular bounce realization. These features act as an additional advantage in 
favour of holographic bounce and reveal the capabilities of the model.

\begin{acknowledgments}
This work is partly supported   by MEXT KAKENHI Grant-in-Aid for 
Scientific Research on Innovative Areas gCosmic Accelerationh No. 15H05890 (S.N.) 
and the JSPS Grant-in-Aid for Scientific Research (C) No. 18K03615 (S.N.), 
and by MINECO (Spain), FIS2016-76363-P, and 
by project 2017 SGR247 (AGAUR, Catalonia) (S.D.O). 
\end{acknowledgments}


\begin{thebibliography}{99}

 
  
  
  
\bibitem{Li:2004rb} 
  M.~Li,
      \href{\doi/10.1016/j.physletb.2004.10.014}{Phys.\ Lett.\ B {\bf 603}, 1 (2004)}
  [\href{\arxiv/hep-th/0403127}{hep-th/0403127}].
   
 

\bibitem{Wang:2016och} 
  S.~Wang, Y.~Wang and M.~Li,
 \href{\doi/10.1016/j.physrep.2017.06.003}{Phys.\ Rept.\  {\bf 696}, 1 (2017)}
 [\href{\arxiv/arXiv:1612.00345}{1612.00345} [astro-ph.CO]].
 
   
\bibitem{tHooft:1993dmi} 
  G.~'t Hooft,
          Salamfest 1993: 0284-296
  [\href{\arxiv/gr-qc/9310026}{gr-qc/9310026}].
 
 
 
\bibitem{Susskind:1994vu} 
  L.~Susskind,
          \href{\doi/10.1063/1.531249}{ J.\ Math.\ Phys.\  {\bf 36}, 6377 (1995)}
  [\href{\arxiv/hep-th/9409089}{hep-th/9409089}].
 
  
\bibitem{Witten:1998qj} 
  E.~Witten,
  \href{\doi/10.4310/ATMP.1998.v2.n2.a2}
  {Adv.\ Theor.\ Math.\ Phys.\  {\bf 2}, 253 (1998)}  
  [\href{\arxiv/hep-th/9802150}{hep-th/9802150}].
 
  
  
\bibitem{Bousso:2002ju} 
  R.~Bousso,
          \href{\doi/10.1103/RevModPhys.74.825}{Rev.\ Mod.\ Phys.\  {\bf 74}, 825 (2002)}
  [\href{\arxiv/hep-th/0203101}{hep-th/0203101}].
 
   
 
    
\bibitem{Kiritsis:2013gia} 
  E.~Kiritsis,
          \href{\doi/10.1088/1475-7516/2013/11/011}{JCAP {\bf 1311}, 011 (2013)}
            [\href{\arxiv/arXiv:1307.5873}{1307.5873} [hep-th]].
  

 
 
  
  
\bibitem{Cohen:1998zx} 
  A.~G.~Cohen, D.~B.~Kaplan and A.~E.~Nelson,
    \href{\doi/10.1103/PhysRevLett.82.4971}{Phys.\ Rev.\ Lett.\  {\bf 82}, 4971 (1999)}
  [\href{\arxiv/hep-th/9803132}{hep-th/9803132}].
 
  
  
  
\bibitem{Horvat:2004vn} 
  R.~Horvat,
       \href{\doi/10.1103/PhysRevD.70.087301}{ Phys.\ Rev.\ D {\bf 70}, 087301 (2004)}
  [\href{\arxiv/astro-ph/0404204}{astro-ph/0404204}].
  
 
  
    
\bibitem{Huang:2004ai} 
  Q.~G.~Huang and M.~Li,
            \href{\doi/10.1088/1475-7516/2004/08/013}{ JCAP {\bf 0408}, 013 (2004)}
  [\href{\arxiv/astro-ph/0404229}{astro-ph/0404229}].
  
 
\bibitem{Pavon:2005yx} 
  D.~Pavon and W.~Zimdahl,
 \href{\doi/10.1016/j.physletb.2005.08.134}{Phys.\ Lett.\ B {\bf 628}, 206 
(2005)}
  [\href{\arxiv/gr-qc/0505020}{gr-qc/0505020}].
  
  
   
\bibitem{Wang:2005jx} 
  B.~Wang, Y.~g.~Gong and E.~Abdalla,
          \href{\doi/10.1016/j.physletb.2005.08.008}{  Phys.\ Lett.\ B {\bf 624}, 141 
(2005)}
  [\href{\arxiv/hep-th/0506069}{hep-th/0506069}].
  
  
   
  
   
 
   
\bibitem{Setare:2008pc} 
  M.~R.~Setare and E.~N.~Saridakis,
         \href{\doi/10.1016/j.physletb.2008.12.026}{Phys.\ Lett.\ B {\bf 671}, 331 (2009)}
  [\href{\arxiv/arXiv:0810.0645}{0810.0645} [hep-th]].
  
    
  
    
\bibitem{Enqvist:2004xv}
K.~Enqvist and M.~S.~Sloth,
           \href{\doi/10.1103/PhysRevLett.93.221302}{Phys.\ Rev.\ Lett.\  {\bf 93} (2004) 
221302}
  [\href{\arxiv/hep-th/0406019}{hep-th/0406019}].

\bibitem{Zhang:2005yz}
X.~Zhang,
  \href{\doi/10.1142/S0218271805007243}{Int.\ J.\ Mod.\ Phys.\ D {\bf 14} (2005) 1597}
  [\href{\arxiv/astro-ph/0504586}{astro-ph/0504586}].
  
  
 

\bibitem{Nojiri:2005pu} 
  S.~Nojiri and S.~D.~Odintsov,
          \href{\doi/10.1007/s10714-006-0301-6}{ Gen.\ Rel.\ Grav.\  {\bf 38}, 1285 
(2006)}
  [\href{\arxiv/hep-th/0506212}{hep-th/0506212}].
   

    
  

\bibitem{Guberina:2005fb}
B.~Guberina, R.~Horvat and H.~Stefancic,
  \href{\doi/10.1088/1475-7516/2005/05/001}{JCAP {\bf 0505} (2005) 001}
  [\href{\arxiv/astro-ph/0503495}{astro-ph/0503495}].
  
 
 
   

 
 

\bibitem{Elizalde:2005ju}
E.~Elizalde, S.~Nojiri, S.~D.~Odintsov and P.~Wang,
  \href{\doi/10.1103/PhysRevD.71.103504}{Phys.\ Rev.\ D {\bf 71} (2005) 103504}
  [\href{\arxiv/hep-th/0502082}{hep-th/0502082}].
  
  
  
    


   

\bibitem{Ito:2004qi}
M.~Ito
  \href{\doi/10.1209/epl/i2005-10151-x}{Europhys.\ Lett.\  {\bf 71} (2005) 712}
  [\href{\arxiv/hep-th/0405281}{hep-th/0405281}].
  
  
   
  
\bibitem{Gong:2004cb}
Y.~g.~Gong, B.~Wang and Y.~Z.~Zhang,
  \href{\doi/10.1103/PhysRevD.72.043510}{Phys.\ Rev.\ D {\bf 72} (2005) 043510}
  [\href{\arxiv/hep-th/0412218}{hep-th/0412218}].
  
   
 
  
\bibitem{Gong:2004fq} 
  Y.~g.~Gong,
   \href{\doi/10.1103/PhysRevD.70.064029}{ Phys.\ Rev.\ D {\bf 70}, 064029 (2004)}
  [\href{\arxiv/hep-th/0404030}{hep-th/0404030}].
  
  
    
 
     
  
   
\bibitem{Setare:2008bb} 
  M.~R.~Setare and E.~C.~Vagenas,
   \href{\doi/10.1016/j.physletb.2008.07.013}{Phys.\ Lett.\ B {\bf 666}, 111 (2008)}
  [\href{\arxiv/arXiv:0801.4478}{0801.4478} [hep-th]].
  
 

  
\bibitem{Saridakis:2007ns} 
  E.~N.~Saridakis,
 \href{\doi/10.1088/1475-7516/2008/04/020}{ JCAP {\bf 0804}, 020 (2008)}
  [\href{\arxiv/arXiv:0712.2672}{0712.2672} [astro-ph]].
  
   
   
  
   

    
\bibitem{Gong:2009dc} 
  Y.~Gong and T.~Li,
     \href{\doi/10.1016/j.physletb.2009.12.040}{Phys.\ Lett.\ B {\bf 683}, 241 (2010)}
  [\href{\arxiv/arXiv:0907.0860}{0907.0860} [hep-th]].
  
   
 
  
    
\bibitem{BouhmadiLopez:2011xi} 
  M.~Bouhmadi-Lopez, A.~Errahmani and T.~Ouali,
       \href{\doi/10.1103/PhysRevD.84.083508}{Phys.\ Rev.\ D {\bf 84}, 083508 (2011)}
  [\href{\arxiv/arXiv:1104.1181}{1104.1181} [astro-ph.CO]].
  
     
    
  
\bibitem{Khurshudyan:2014axa} 
  M.~Khurshudyan, J.~Sadeghi, R.~Myrzakulov, A.~Pasqua and H.~Farahani,
     \href{\doi/10.1155/2014/878092}{Adv.\ High Energy Phys.\  {\bf 2014}, 878092 
(2014)}
  [\href{\arxiv/arXiv:1404.2141}{1404.2141} [gr-qc]].
  
    
     
   
    
    
\bibitem{Pasqua:2016wrm} 
  A.~Pasqua, S.~Chattopadhyay, K.~A.~Assaf and I.~G.~Salako,
 \href{\doi/10.1140/epjp/i2016-16182-7}{Eur.\ Phys.\ J.\ Plus {\bf 131}, no. 6, 182 
(2016)}.

  

    
\bibitem{Jawad:2016tne} 
  A.~Jawad, N.~Azhar and S.~Rani,
   \href{\doi/10.1142/S0218271817500407}{Int.\ J.\ Mod.\ Phys.\ D {\bf 26}, no. 04, 
1750040 (2016)}.
 
\bibitem{Pourhassan:2017cba} 
  B.~Pourhassan, A.~Bonilla, M.~Faizal and E.~M.~C.~Abreu,
  [\href{\arxiv/arXiv:1704.03281}{1704.03281} [hep-th]].
  
\bibitem{Paul:2019hys} 
  T.~Paul,
    [\href{\arxiv/arXiv:1905.13033}{1905.13033} [gr-qc]].
    
 
  
  
 
\bibitem{Zhang:2005hs} 
  X.~Zhang and F.~Q.~Wu,
       \href{\doi/10.1103/PhysRevD.72.043524}{Phys.\ Rev.\ D {\bf 72}, 043524 (2005)}
  [\href{\arxiv/astro-ph/0506310}{astro-ph/0506310}].
  
     
  
\bibitem{Li:2009bn} 
  M.~Li, X.~D.~Li, S.~Wang and X.~Zhang,
         \href{\doi/10.1088/1475-7516/2009/06/036}{JCAP {\bf 0906}, 036 (2009)}
  [\href{\arxiv/arXiv:0904.0928}{0904.0928} [astro-ph.CO]].
  
      
     
\bibitem{Lu:2009iv} 
  J.~Lu, E.~N.~Saridakis, M.~R.~Setare and L.~Xu,
         \href{\doi/10.1088/1475-7516/2010/03/031}{JCAP {\bf 1003}, 031 (2010)}
  [\href{\arxiv/arXiv:0912.0923}{0912.0923} [astro-ph.CO]].
  
    
 
\bibitem{Huang:2004wt}
Q.~G.~Huang and Y.~G.~Gong,
         \href{\doi/10.1088/1475-7516/2004/08/006}{JCAP {\bf 0408} (2004) 006}
  [\href{\arxiv/astro-ph/0403590}{astro-ph/0403590}].
  
  
  
   




\bibitem{Wang:2004nqa}
B.~Wang, E.~Abdalla and R.~K.~Su,
     \href{\doi/10.1016/j.physletb.2005.02.026}{Phys.\ Lett.\ B {\bf 611} (2005) 21}
  [\href{\arxiv/hep-th/0404057}{hep-th/0404057}].
  
  
  

\bibitem{Nojiri:2019kkp} 
  S.~Nojiri, S.~D.~Odintsov and E.~N.~Saridakis,
     [\href{http://xxx.lanl.gov/abs/1904.01345}
  {{\tt arXiv:1904.01345 [gr-qc]}}].
 
 
 
\bibitem{Novello:2008ra}
  M.~Novello and S.~E.~P.~Bergliaffa,
  \href{\doi/10.1016/j.physrep.2008.04.006}   
  {Phys.\ Rept.\  {\bf 463}, 127 (2008)}
  [\href{http://xxx.lanl.gov/abs/0802.1634}
  {{\tt arXiv:0802.1634 [astro-ph]}}].

  
  \bibitem{Cai:2012va} 
  Y.~F.~Cai, D.~A.~Easson and R.~Brandenberger,
     \href{\doi/10.1088/1475-7516/2012/08/020}{JCAP {\bf 1208}, 020 (2012)}
  [\href{http://xxx.lanl.gov/abs/1206.2382}
  {{\tt arXiv:1206.2382 [hep-th]}}].

  
  
\bibitem{Cai:2014bea}
  Y.~F.~Cai,
  \href{\doi/10.1007/s11433-014-5512-3}
  {Sci.\ China Phys.\ Mech.\ Astron.\  {\bf 57}, 1414 (2014)}
  [\href{http://xxx.lanl.gov/abs/1405.1369}
  {{\tt arXiv:1405.1369 [hep-th]}}].
   
  
\bibitem{Brandenberger:2012zb} 
  R.~H.~Brandenberger,
  [\href{http://xxx.lanl.gov/abs/1206.4196}
  {{\tt arXiv:1206.4196 [astro-ph.CO]}}].

  
  
  
 
\bibitem{Brandenberger:2016vhg}
  R.~Brandenberger and P.~Peter,
  \href{\doi/10.1007/s10701-016-0057-0}
  {Found.\ Phys.\ (2017)}
  [\href{http://xxx.lanl.gov/abs/1603.05834}
  {{\tt arXiv:1603.05834 [hep-th]}}].

\bibitem{Cai:2016hea} 
  Y.~F.~Cai, A.~Marciano, D.~G.~Wang and E.~Wilson-Ewing,
  \href{\doi/10.3390/universe3010001}
  {Universe {\bf 3}, no. 1, 1 (2016)}
  [\href{http://xxx.lanl.gov/abs/1610.00938}
  {{\tt arXiv:1610.00938 [astro-ph.CO]}}].
  
  
\bibitem{deHaro:2015wda}
  J.~de Haro and Y.~F.~Cai,
       \href{\doi/10.1007/s10714-015-1936-y}{Gen.\ Rel.\ Grav.\  {\bf 47} (2015) no.8,  
95}
  [\href{\arxiv/arXiv:1502.03230}{1502.03230} [gr-qc]].
  
   
    
  

\bibitem{Nojiri:2006ri}
  S.~i.~Nojiri and S.~D.~Odintsov,
  eConf C {\bf 0602061}, 06 (2006)
  \href{\doi/10.1142/S0219887807001928}
  {Int.\ J.\ Geom.\ Meth.\ Mod.\ Phys.\  {\bf 4}, 115 (2007)}
  [\href{http://xxx.lanl.gov/abs/hep-th/0601213}
  {{\tt arXiv:hep-th/0601213}}].

\bibitem{Capozziello:2011et}
  S.~Capozziello and M.~De Laurentis,
  \href{\doi/10.1016/j.physrep.2011.09.003}
  {Phys.\ Rept.\  {\bf 509}, 167 (2011)}
  [\href{http://xxx.lanl.gov/abs/1108.6266}
  {{\tt arXiv:1108.6266 [gr-qc]}}].

\bibitem{Cai:2015emx} 
  Y.~F.~Cai, S.~Capozziello, M.~De Laurentis and E.~N.~Saridakis,
  \href{\doi/10.1088/0034-4885/79/10/106901}
  {Rept.\ Prog.\ Phys.\  {\bf 79}, no. 10, 106901 (2016)}
  [\href{http://xxx.lanl.gov/abs/1511.07586}
  {{\tt arXiv:1511.07586 [gr-qc]}}].
  
  
  
  

\bibitem{Veneziano:1991ek}
  G.~Veneziano,
  \href{\doi/10.1016/0370-2693(91)90055-U}
  {Phys.\ Lett.\  B {\bf 265}, 287 (1991)}.

\bibitem{Khoury:2001wf}
  J.~Khoury, B.~A.~Ovrut, P.~J.~Steinhardt and N.~Turok,
  \href{\doi/10.1103/PhysRevD.64.123522}
  {Phys.\ Rev.\  D {\bf 64}, 123522 (2001)}
  [\href{http://xxx.lanl.gov/abs/hep-th/0103239}
  {{\tt arXiv:hep-th/0103239}}].

  

\bibitem{Bamba:2013fha}
  K.~Bamba, A.~N.~Makarenko, A.~N.~Myagky, S.~Nojiri and S.~D.~Odintsov,
  \href{\doi/10.1088/1475-7516/2014/01/008}
  {JCAP {\bf 1401} (2014) 008}
  [\href{http://xxx.lanl.gov/abs/1309.3748}
  {{\tt arXiv:1309.3748 [hep-th]}}].

 

\bibitem{Pavlovic:2017umo} 
  P.~Pavlovic and M.~Sossich,
  \href{\doi/10.1103/PhysRevD.95.103519}
  {Phys.\ Rev.\ D {\bf 95}, no. 10, 103519 (2017)}
  [\href{http://xxx.lanl.gov/abs/1701.03657}
  {{\tt arXiv:1701.03657 [gr-qc]}}].

\bibitem{Cai:2011tc}
  Y.~F.~Cai, S.~-H.~Chen, J.~B.~Dent, S.~Dutta and E.~N.~Saridakis,
  \href{\doi/10.1088/0264-9381/28/21/215011}
  {Class.\ Quant.\ Grav.\  {\bf 28}, 215011 (2011)}
  [\href{http://xxx.lanl.gov/abs/1104.4349}
  {{\tt arXiv:1104.4349 [astro-ph.CO]}}].

 

 
 
 

\bibitem{Bojowald:2001xe}
  M.~Bojowald,
  \href{\doi/10.1103/PhysRevLett.86.5227}
  {Phys.\ Rev.\ Lett.\  {\bf 86}, 5227 (2001)}
  [\href{http://xxx.lanl.gov/abs/gr-qc/0102069}
  {{\tt arXiv:gr-qc/0102069}}].

\bibitem{Cai:2014zga} 
  Y.~F.~Cai and E.~Wilson-Ewing,
  \href{\doi/10.1088/1475-7516/2014/03/026}
  {JCAP {\bf 1403}, 026 (2014)}
  [\href{http://xxx.lanl.gov/abs/1402.3009}
  {{\tt arXiv:1402.3009 [gr-qc]}}].

\bibitem{Odintsov:2015uca} 
  S.~D.~Odintsov, V.~K.~Oikonomou and E.~N.~Saridakis,
  \href{\doi/10.1016/j.aop.2015.08.021}
  {Annals Phys.\  {\bf 363}, 141 (2015)}
  [\href{http://xxx.lanl.gov/abs/1501.06591}
  {{\tt arXiv:1501.06591 [gr-qc]}}].

\bibitem{Cai:2010zma}
  Y.~F.~Cai and E.~N.~Saridakis,
  \href{\doi/10.1088/0264-9381/28/3/035010}
  {Class.\ Quant.\ Grav.\  {\bf 28}, 035010 (2011)}
  [\href{http://xxx.lanl.gov/abs/1007.3204}
  {{\tt arXiv:1007.3204 [astro-ph.CO]}}].
  
  
\bibitem{Minas:2019urp} 
  G.~Minas, E.~N.~Saridakis, P.~C.~Stavrinos and A.~Triantafyllopoulos,
    \href{\doi/10.3390/universe5030074}
  {Universe {\bf 5}, 74 (2019)}
  [\href{http://xxx.lanl.gov/abs/1902.06558}
  {{\tt arXiv:1902.06558 [gr-qc]}}].
 
 
  
  
  

\bibitem{Cai:2007qw}
  Y.~F.~Cai, T.~Qiu, Y.~S.~Piao, M.~Li and X.~Zhang,
  \href{\doi/10.1088/1126-6708/2007/10/071}
  {JHEP {\bf 0710}, 071 (2007)}
  [\href{http://xxx.lanl.gov/abs/0704.1090}
  {{\tt arXiv:0704.1090 [gr-qc]}}].

 

\bibitem{Cai:2009zp}
  Y.~F.~Cai, E.~N.~Saridakis, M.~R.~Setare and J.~Q.~Xia,
  \href{\doi/10.1016/j.physrep.2010.04.001}
  {Phys.\ Rept.\  {\bf 493}, 1 (2010)}
  [\href{\arxiv/arXiv:0909.2776}
  {{\tt arXiv:0909.2776 [hep-th]}}].
  
  
 
  
  
  
  
  
  
  
  
  
  
  
  
  
  
  
  
  
    
     
\bibitem{Hsu:2004ri} 
  S.~D.~H.~Hsu,
     \href{\doi/10.1016/j.physletb.2004.05.020}{Phys.\ Lett.\ B {\bf 594}, 13 (2004)}
  [\href{\arxiv/hep-th/0403052}{hep-th/0403052}].
  
  
  
\bibitem{Cai:2007us} 
  R.~G.~Cai,
    \href{\doi/10.1016/j.physletb.2007.09.061}{Phys.\ Lett.\ B {\bf 657}, 228 (2007)}
  [\href{\arxiv/arXiv:0707.4049}{0707.4049} [hep-th]].
   

\bibitem{Wei:2007ty} 
  H.~Wei and R.~G.~Cai,
   \href{\doi/10.1016/j.physletb.2007.12.030}{Phys.\ Lett.\ B {\bf 660}, 113 (2008)}
  [\href{\arxiv/arXiv:0708.0884}{0708.0884} [astro-ph]].
  
 
  
   
    

    
\bibitem{Gao:2007ep} 
  C.~Gao, X.~Chen and Y.~G.~Shen,
   \href{\doi/10.1103/PhysRevD.79.043511}{Phys.\ Rev.\ D {\bf 79}, 043511 (2009)}
 [\href{\arxiv/arXiv:0712.1394}{0712.1394} [astro-ph]].
       



\bibitem{Saridakis:2017rdo} 
  E.~N.~Saridakis,
     \href{\doi/10.1103/PhysRevD.97.064035}{  Phys.\ Rev.\ D {\bf 97}, no. 6, 064035 
(2018)}
 [\href{\arxiv/arXiv:1707.09331}{1707.09331} [gr-qc]].
  
  
   
  
\bibitem{Nojiri:2017opc} 
  S.~Nojiri and S.~D.~Odintsov,
       \href{\doi/10.1140/epjc/s10052-017-5097-x}{Eur.\ Phys.\ J.\ C {\bf 77}, no. 8, 
528 (2017)}
 [\href{\arxiv/arXiv:1703.06372}{1703.06372} [hep-th]].
  
  
   
  
\bibitem{Nojiri:2004pf}
S.~Nojiri and S.~D.~Odintsov,
       \href{\doi/10.1103/PhysRevD.70.103522}{Phys.\ Rev.\ D {\bf 70} (2004) 103522}
 [\href{\arxiv/hep-th/0408170}{hep-th/0408170}].
  
   
  
 
   
  
\bibitem{Battarra:2014tga} 
  L.~Battarra, M.~Koehn, J.~L.~Lehners and B.~A.~Ovrut,
         \href{\doi/10.1088/1475-7516/2014/07/007}{JCAP {\bf 1407}, 007 (2014)}
 [\href{\arxiv/arXiv:1404.5067}{1404.5067} [hep-th]].
  
   


\bibitem{Quintin:2015rta} 
  J.~Quintin, Z.~Sherkatghanad, Y.~F.~Cai and R.~H.~Brandenberger,
         \href{\doi/10.1103/PhysRevD.92.063532}{Phys.\ Rev.\ D {\bf 92}, no. 6, 063532 
(2015)}
 [\href{\arxiv/arXiv:1508.04141}{1508.04141} [hep-th]].
  
  
\bibitem{Allen:2004vz} 
  L.~E.~Allen and D.~Wands,
           \href{\doi/10.1103/PhysRevD.70.063515}{Phys.\ Rev.\ D {\bf 70}, 063515 
(2004)}
 [\href{\arxiv/arXiv:astro-ph/0404441}{astro-ph/0404441}].
  
  
  
  

   
   
\bibitem{Battefeld:2004jh} 
  T.~J.~Battefeld and D.~A.~Easson,
           \href{10.1103/PhysRevD.70.103516}{Phys.\ Rev.\ D {\bf 70}, 103516 (2004)}
   [\href{\arxiv/arXiv:hep-th/0408154}{hep-th/0408154}].
   
 
\bibitem{Kim:2008pi} 
  K.~Y.~Kim, H.~W.~Lee and Y.~S.~Myung,
           \href{\doi/10.1142/S0217732309030643}{Mod.\ Phys.\ Lett.\ A {\bf 24}, 1267 
(2009)}
   [\href{\arxiv/arXiv:0805.3941}{0805.3941}].
   
 
  
\bibitem{Mehrabi:2015kta} 
  A.~Mehrabi, S.~Basilakos, M.~Malekjani and Z.~Davari,
             \href{\doi/10.1103/PhysRevD.92.123513}{Phys.\ Rev.\ D {\bf 92}, no. 12, 
123513 (2015)}
   [\href{\arxiv/arXiv:1510.03996}{1510.03996}].
  
\bibitem{Li:2016xjb} 
  Y.~B.~Li, J.~Quintin, D.~G.~Wang and Y.~F.~Cai,
  \href{\doi/10.1088/1475-7516/2017/03/031}
  {JCAP {\bf 1703}, no. 03, 031 (2017)}
  [\href{http://xxx.lanl.gov/abs/1612.02036}
  {{\tt arXiv:1612.02036 [hep-th]}}].

\bibitem{Belinsky:1970ew} 
  V.~A.~Belinsky, I.~M.~Khalatnikov and E.~M.~Lifshitz,
 {\it{Oscillatory approach to a singular point in the relativistic cosmology}},
  Adv.\ Phys.\  {\bf 19}, 525 (1970).



  
\bibitem{Qiu:2013eoa} 
  T.~Qiu, X.~Gao and E.~N.~Saridakis,
 {\it{Towards anisotropy-free and nonsingular bounce cosmology with scale-invariant 
perturbations}},
  Phys.\ Rev.\ D {\bf 88}, no. 4, 043525 (2013)
  [\href{http://xxx.lanl.gov/abs/1303.2372}
{{\tt arXiv:1303.2372}}].




\bibitem{Cai:2013vm} 
  Y.~F.~Cai, R.~Brandenberger and P.~Peter,
 {\it{Anisotropy in a Nonsingular Bounce}},
  Class.\ Quant.\ Grav.\  {\bf 30}, 075019 (2013),
    [\href{http://xxx.lanl.gov/abs/1301.4703}
{{\tt arXiv:1301.4703}}].


 
 
  
 
 
\bibitem{Cai:2011ci} 
  Y.~F.~Cai, R.~Brandenberger and X.~Zhang,
           \href{\doi/10.1016/j.physletb.2011.07.074}{Phys.\ Lett.\ B {\bf 703}, 25 
(2011)}
   [\href{\arxiv/arXiv:1105.4286}{1105.4286} [hep-th]].

 
  
 
   
  
\bibitem{Cai:2016sdu} 
  Y.~F.~Cai, S.~Lin, J.~Liu and J.~R.~Sun,
 [\href{\arxiv/arXiv:1612.04377}{1612.04377} [hep-th]].
   

 
\bibitem{Cai:2016lqa} 
  Y.~F.~Cai, S.~Lin, J.~Liu and J.~R.~Sun,
   [\href{\arxiv/arXiv:1612.04394}{1612.04394} [hep-th]].
   
 

     


\bibitem{Nojiri:2017ncd}
S.~Nojiri, S.~D.~Odintsov and V.~K.~Oikonomou,
     \href{\doi/10.1016/j.physrep.2017.06.001}{Phys.\ Rept.\  {\bf 692} (2017) 1}
 [\href{\arxiv/arXiv:1705.11098}{1705.11098} [gr-qc]].
  
  
   



\bibitem{Nojiri:2010wj}
S.~Nojiri and S.~D.~Odintsov,
     \href{\doi/10.1016/j.physrep.2011.04.001}{Phys.\ Rept.\  {\bf 505} (2011) 59}
 [\href{\arxiv/arXiv:1011.0544}{1011.0544} [gr-qc]].
   



 
 

    
\bibitem{Olmo:2011uz}
G.~J.~Olmo,
     \href{\doi/10.1142/S0218271811018925}{Int.\ J.\ Mod.\ Phys.\ D {\bf 20} (2011) 413}
 [\href{\arxiv/arXiv:1101.3864}{1101.3864} [gr-qc]].
   

 
     
  
 
   
    

 

\bibitem{Capozziello:2010zz}
V.~Faraoni and S.~Capozziello,
  \href{\doi/10.1007/978-94-007-0165-6}{Fundam.\ Theor.\ Phys.\  {\bf 170} (2010)}.
 
   
 
 

\bibitem{delaCruzDombriz:2012xy}
A.~de la Cruz-Dombriz and D.~Saez-Gomez,
         \href{\doi/10.3390/e14091717}{ Entropy {\bf 14} (2012) 1717}
   [\href{\arxiv/arXiv:1207.2663}{1207.2663} [gr-qc]].
   

   

    
\end{thebibliography}
\end{document}